\documentclass[%
 reprint,
 amsmath,amssymb,
 aps, superscriptaddress
]{revtex4-1}

\usepackage{graphicx}
\usepackage{graphicx}
\usepackage{dcolumn}
\usepackage{bm}
\usepackage{soul}
\usepackage{color}
\usepackage{epstopdf}
\usepackage[version=3]{mhchem}
\usepackage{lipsum}
\usepackage[outercaption]{sidecap}
\usepackage{floatrow}

\begin{document}

\preprint{APS/123-QED}

\title{Multilayered plasmonic nanostructures for solar energy harvesting}

\author{Anh D. Phan}
\affiliation{Department of Physics, University of Illinois, 1110 West Green St, Urbana, Illinois 61801, USA}
\affiliation{Institute of Physics, Vietnam Academy of Science and Technology, 10 Dao Tan, Ba Dinh, Hanoi 10000, Vietnam}
\email{adphan35@gmail.com}
\author{Nam B. Le}
\affiliation{School of Engineering Physics, Hanoi University of Science and Technology, 1 Dai Co Viet, Hanoi, Vietnam}
\email{nble@mail.usf.edu}
\author{Nghiem T. H. Lien}
\affiliation{Institute of Physics, Vietnam Academy of Science and Technology, 10 Dao Tan, Ba Dinh, Hanoi 10000, Vietnam}
\author{Katsunori Wakabayashi}
\affiliation{Department of Nanotechnology for Sustainable Energy, School of Science and Technology, Kwansei Gakuin University, Sanda, Hyogo 669-1337, Japan}
\date{\today}

\begin{abstract}
Optical properties of core-shell-shell Au@\ce{SiO_2}@Au nanostructures and their solar energy harvesting applications are theoretically investigated using Mie theory and heat transfer equations. The theoretical analysis associated with size-dependent modification of the bulk gold dielectric function agrees well with previous experimental results. We use the appropriate absorption cross section to determine the solar energy absorption efficiency of the nano-heterostructures, which is strongly structure-dependent, and to predict the time-dependent temperature increase of the nanoshell solution under simulated solar irradiation. Comparisons to prior temperature measurements and theoretical evaluation of the solar power conversion efficiency are discussed to provide new insights into underlying mechanisms. Our approach would accelerate materials and structure testing in solar energy harvesting.
\end{abstract}

\pacs{Valid PACS appear here}
\maketitle

\section{Introduction}
Multilayer metal-based nanoshells have been the subject of much recent research interest because of significant near-field enhancement and peculiar properties compared with those of single-component counterparts \cite{1,9,18,19,20,21,27}. A composite material not only possesses original characteristics of its individual components, but also new extraordinary attributes caused by coupling between them. Strong light-matter interactions originate from excitations of the collective oscillations of conduction electrons at metallic surfaces, so-called surface plasmon resonances \cite{1,18,19}. Optical features of metal nanoshells remarkably depend on their size, shape, and ambient medium \cite{1}. The extraordinary light manipulation abilities of plasmonic heterostructures have been exploited for a wide range of applications such as spectroscopy \cite{28,29}, medical treatment \cite{3,27}, high-efficiency photovoltaics \cite{30} and solar cells \cite{24,25,26}. Among these applications, improving solar cell efficiency is an essential path to deal with the energy crisis when supplies of conventional energy resources are exhausted. 

Recently, hybrid nanocomposites composed of gold and silica have been intensively investigated for light-to-heat conversion \cite{24,26,13,17}. While gold nanoparticles have fascinating properties derived from their localized surface plasmon resonances \cite{1}, the use of silica prevents electron transfer between materials, protects interior metal from corrosion, and red-shifts plasmon peaks. Additionally, inserting a dielectric layer between two gold layers of the nanostructures creates an additional resonance, enhances absorption intensity, and broadens the absorption band \cite{9,27,28,6}. Properly synthesizing the size of nanostructures can optimize solar energy harvesting and effectively release heat. 





\begin{figure}[htp]
\includegraphics[width=5cm]{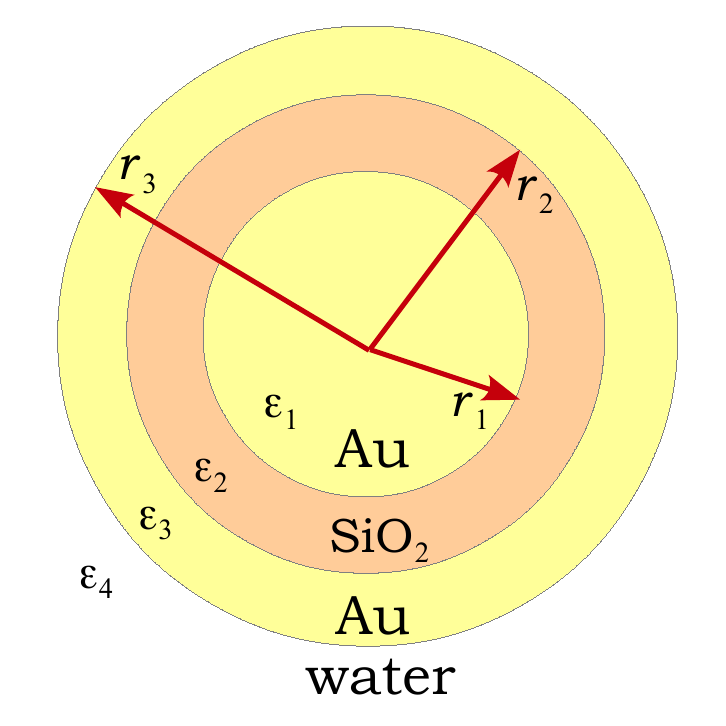}
\caption{\label{fig:1}(Color online) Schematic illustration of gold nanomatryoshkas.} 
\end{figure}

In this work, we propose a theoretical approach to predict time-dependent temperature rise of gold-silica-gold multilayer nanoshells (as depicted in Fig.\ref{fig:1}) dispersed in water under the illumination of simulated solar light at 80 mW/\ce{cm^2}. We carefully analyze structurally dependent effects of gold nanomatryoshkas on extinction spectra by comparing experiments with Mie scattering calculations. The corresponding absorption cross sections of the hybrid nanostructures are used to estimate the heat source density and the solar energy absorption efficiency by using a proposed figure of merit \cite{22}. Then thermal fields generated by the heat source are analytically formulated. Our approach could quantitatively predict future experiments.

\section{Theoretical background}
\subsection{Mie scattering theory}
Mie theory has been used widely and successfully to calculate the absorption, scattering, and extinction spectra of multilayered nanocomposites embedded in a homogeneous and isotropic medium \cite{1,3,16}. The approach shows a good agreement with experimental results. Accurate quantitative predictions of absorption cross sections can be used to investigate spatial and temporal temperature variations when a material is illuminated by laser \cite{3,4} and solar radiation. Analytical expressions of theoretical optical cross-sections predicted by Mie scattering calculations are \cite{1}

\begin{eqnarray}
Q_{ext} &=& \frac{2\pi}{k_4^2}\sum_{n=1}^{\infty}(2n+1){Re}\left(a_n + b_n \right), \nonumber\\
Q_{scat} &=& \frac{2\pi}{k_4^2}\sum_{n=1}^{\infty}(2n+1)\left(\left | a_n \right|^2  + \left | b_n \right |^2 \right),\nonumber\\
Q_{abs} &=& Q_{ext}  - Q_{scat},
\label{eq:1}
\end{eqnarray}

where

\begin{widetext}
\begin{eqnarray}
a_n=\frac{\Psi_n(x_3)\left[\Psi'_n(m_3x_3)+S_n^2\chi_n'(m_3x_3) \right]-m_3\Psi'_n(x_3)\left[\Psi_n(m_3x_3)+ S_n^2\chi_n(m_3x_3)\right]}{\xi_n(x_3)\left[\Psi'_n(m_3x_3)+S_n^2\chi_n'(m_3x_3) \right]-m_3\xi'_n(x_3)\left[\Psi_n(m_3x_3)+ S_n^2\chi_n(m_3x_3)\right]},\\
b_n=\frac{m_3\Psi_n(x_3)\left[\Psi'_n(m_3x_3)+T_n^2\chi_n'(m_3x_3) \right]-\Psi'_n(x_3)\left[\Psi_n(m_3x_3)+ T_n^2\chi_n(m_3x_3)\right]}{m_3\xi_n(x_3)\left[\Psi'_n(m_3x_3)+T_n^2\chi_n'(m_3x_3) \right]-\xi'_n(x_3)\left[\Psi_n(m_3x_3)+ T_n^2\chi_n(m_3x_3)\right]},\\
S_n^s=\frac{\Psi_n(x_s)\left[\Psi'_n(m_sx_s)+S_n^{s-1}\chi_n'(m_sx_s) \right]-m_s\Psi'_n(x_s)\left[\Psi_n(m_sx_s)+ S_n^{s-1}\chi_n(m_sx_s)\right]}{\xi_n(x_s)\left[\Psi'_n(m_sx_s)+S_n^{s-1}\chi_n'(m_sx_s) \right]-m_s\xi'_n(x_s)\left[\Psi_n(m_sx_s)+ S_n^{s-1}\chi_n(m_sx_s)\right]},\\
T_n^s=\frac{m_s\Psi_n(x_s)\left[\Psi'_n(m_sx_s)+T_n^{s-1}\chi_n'(m_sx_s) \right]-\Psi'_n(x_s)\left[\Psi_n(m_sx_s)+ T_n^{s-1}\chi_n(m_sx_s)\right]}{m_s\xi_n(x_s)\left[\Psi'_n(m_sx_s)+T_n^{s-1}\chi_n'(m_sx_s) \right]-\xi'_n(x_s)\left[\Psi_n(m_sx_s)+ T_n^{s-1}\chi_n(m_sx_s)\right]},
\label{eq:2}
\end{eqnarray}
\end{widetext}
where $S_n^0=T_n^0=0$, $x_s=k_{s+1}r_s$, $m_s=\sqrt{\varepsilon_s/\varepsilon_{s+1}}$, $\varepsilon_s$ is the dielectric function in the $s$-th layer, $k_s=2\pi\sqrt{\varepsilon_s}/\lambda$ is the wavenumber, $\lambda$ is the wavelength of incident light in vacuum, $r_s$ is a distance from a center of the multilayered nanostructures to the $s$-th interface, $\Psi(x)=xj_n(x)$, $\chi_n(x)=xy_n(x)$, and $\xi_n(x) = xh_n^{(1)}(x)$ are Riccati-Bessel, Riccati-Neumann and Riccati-Hankel functions, respectively. The $s=1, 2,$ and 3 indicates media in the gold nanomatryoshkas. $j_n(x)$ is the spherical Bessel function of the first kind, $y_n(x)$ is the spherical Neumann function, and $h_n^{(1)}(x)$ is the spherical Hankel function of the first kind. The Mie scattering coefficients $a_n$ and $b_n$ correspond to transverse magnetic (TM) and transverse electric (TE) modes, respectively. In our calculations, the dielectric function of silica $\varepsilon_2$ is approximately 2.04, the dielectric permittivity of ambient medium $\varepsilon_4 = 1.77$, while the dielectric function of gold ($\varepsilon_{1}(\omega)=\varepsilon_{3}(\omega)$) is described by the Lorentz-Drude model with several oscillators \cite{2,3,4,5}
\begin{eqnarray}
\varepsilon_{1,3}(\omega) = 1- \frac{f\omega_p^2}{\omega^2-i\omega\Gamma_0} + \sum_{j=1}^{5}\frac{f_j\omega_p^2}{\omega_j^2-\omega^2+i\omega\Gamma_j},
\label{eq:3}
\end{eqnarray}
where $f_0$ and $f_j$ are the oscillator strengths, $\omega_p$ is the plasma frequency for gold, and $\Gamma_0$ and $\Gamma_j$ are the damping parameters. All parameters in this model come from Ref. \cite{2}. To consider size effects on optical properties of the multilayered nanostructures, the damping parameter in Eq.(\ref{eq:3}) is modified by $\Gamma_0 \equiv \Gamma_0+Av_F/d_s$, where $v_F$ is the gold Fermi velocity, $d_s$ is the thickness of the $s$-th gold layer, and $A$ is the parameter characterizing the scattering processes. {\color{red} The parameter $A$ ranging from 0 to 3 \cite{1,3,16,4} is empirically selected to achieve the best fitting curve for optical measurements.}{\color{red}The size-dependent correction of the damping parameter loses validity for the gold thickness $d_s$ below 1-2 nm \cite{34,35,36}.} 

\subsection{Solar energy conversion}
In this subsection, we propose a theoretical approach for photothermal effects under the same conditions with the sunlight heating experiments in Ref. \cite{22}. After illuminating a volume $V_0$ of an aqueous solution of gold nanoshells by a sunlight simulator (XES-40S1, San-Ei Electric), electromagnetic fields excite the surface localized resonance, and the nanostructures are heated through a light-to-heat conversion mechanism. Here $V_0 = 20$ $ml$ identical to Ref. \cite{22}. One can assume that the particles are randomly dispersed in a spherical region of radius $R = (3V_0/4\pi)^{1/3}$ and the simulated solar energy is perfectly converted to heat energy. The heat source density caused by absorbed energy on the nanoparticles (NP) from the solar simulator is 
\begin{eqnarray}
P_{abs}^{NP} = N\int_{300}^{1100} Q_{abs}(\lambda)E_\lambda d\lambda,
\label{eq:10}
\end{eqnarray}
where $E_\lambda$ is the solar spectral irradiance of AM1.5 G and $N$ is the number of particles per unit volume in the solution. Note that the electromagnetic radiation of the XES-40S1 solar simulator is in the wavelength range between 300 nm and 1100 nm. While the absorbed heat flux because of pure water is given by
\begin{eqnarray}
P_{abs}^{w} &=& \int_{300}^{1100} E_\lambda d\lambda \left(1-\left\langle e^{-2\alpha(\lambda)R} \right\rangle \right),\nonumber\\
\left\langle e^{-2\alpha(\lambda)R} \right\rangle &=& \frac{1}{1100-300}\int_{300}^{1100} e^{-2\alpha(\lambda)R}d\lambda,
\label{eq:11}
\end{eqnarray}
where $\alpha(\lambda)$ is the absorption coefficient \cite{23}. 

Having assumed that the temperature gradient is homogeneous during the heating process, temperature variation of samples $\Delta T(t)$ induced by the photothermal effect of gold composites is theoretically described by the heat energy balance equation \cite{3,4} 
\begin{eqnarray}
\rho c V_0\frac{\partial \Delta T}{\partial t} = P_{abs}^{NP}V_0 + P_{abs}^{w}4\pi R^2,
\label{eq:7}
\end{eqnarray}
where $\rho = 1000$ $kg/m^3$ is the mass density of water and $c=4200$ $J/kg/K$ is the specific heat of water. The presence of the heat generation exists only within the localized domain of the nanoshells since there are no nanoparticles outside the region. 

\section{Results and discussions}
\begin{figure}[h]
\includegraphics[width=8cm]{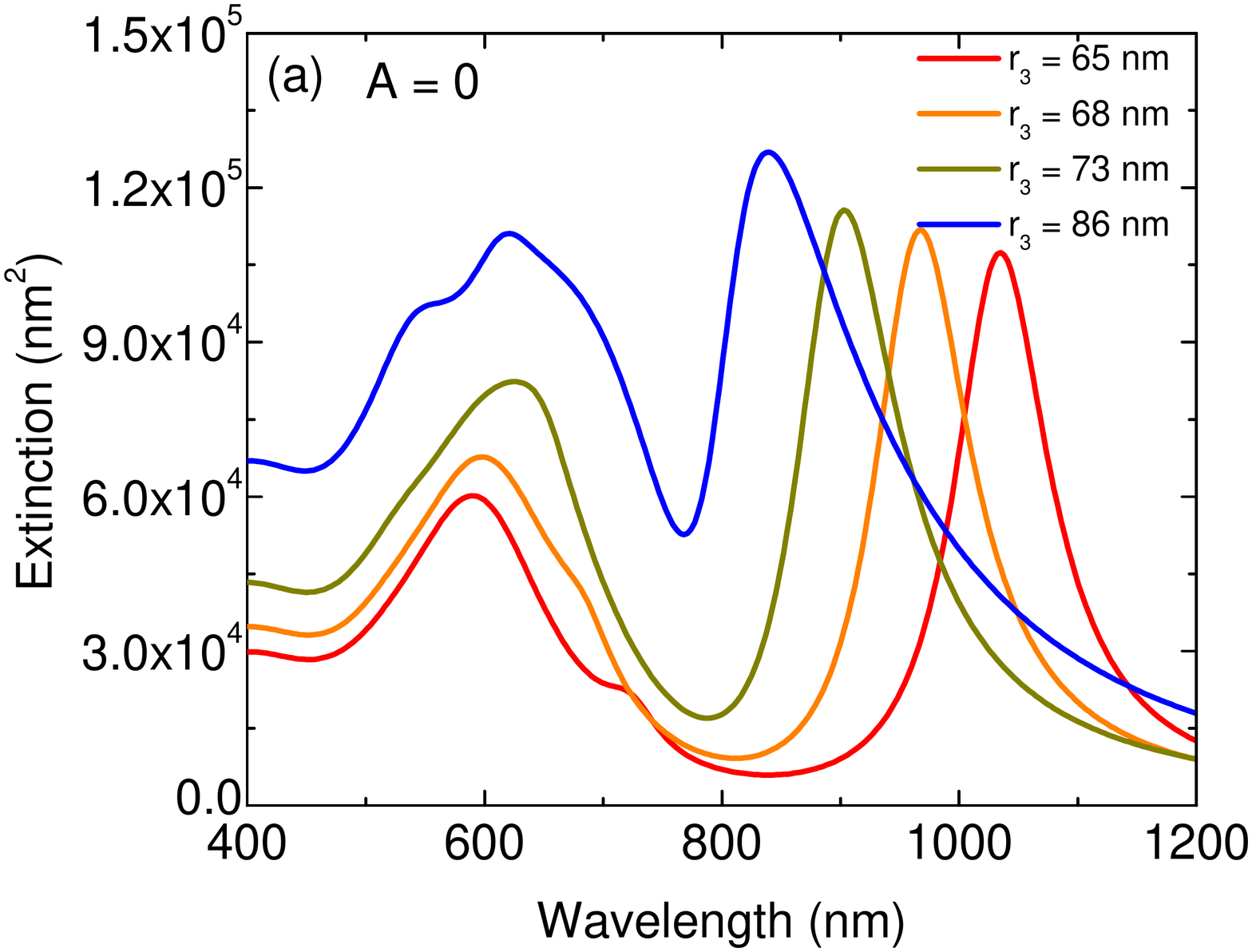}\\
\includegraphics[width=8cm]{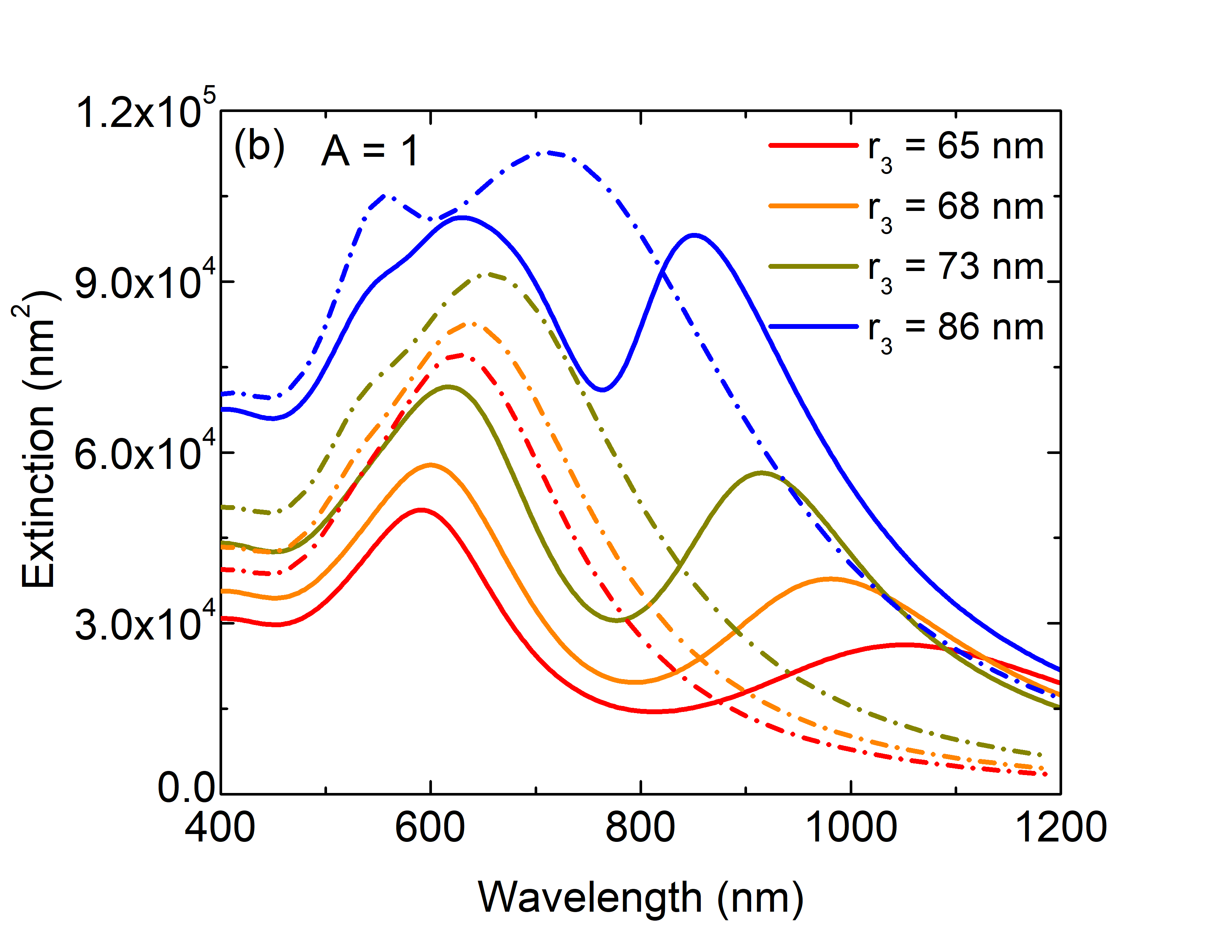}
\caption{\label{fig:2}(Color online) Theoretical extinction cross section of gold nanomatryoshkas of radius $r_1=40$ nm and $r_2 = 55$ nm at several values of radius $r_3$, computed using Mie theory associated with the different size effect parameters: (a) $A = 0$ and (b) $A = 1$ to mimic experiments in Ref.\cite{9}. The dashed-dotted curves are the analogous results for a single component gold nanoparticle having radius $r_3$.}
\end{figure}

Figure \ref{fig:2} shows the theoretical extinction spectra of Au-\ce{SiO_2}-Au nanoshells of outer radius $r_3$ = 65, 68, 73 and 86 nm with several values of the parameter $A$. Two resonances at 592 nm and 940 nm are attributed to collective motions of electrons on the surface of the gold core and outer shell, respectively, in the nanostructure. Increasing $r_3$ induces a blue shift of the second optical peak because of a weak coupling between the gold inner and outer surfaces. For $A = 0$, size effects are not captured on the spectra, and our results are almost identical to the previous theoretical analysis in Ref. \cite{9} using Mie theory and bulk Johnson and Christy dielectric function for gold \cite{7}. As seen in Fig. \ref{fig:2}a, the intensity of the second peak is higher than the first peak, whereas the experiment in Ref.\cite{9} shows the opposite. Therefore, the finite-size effects have to be taken into account when investigating optical properties of nanoscale composite materials. Numerical results in Fig. \ref{fig:2}b indicate that increasing parameter $A$ does not significantly change the two resonance positions but the increase lowers the extinction cross section of the peak in the near-infrared region. Interestingly, for $A = 1$, the plasmonic features of the theoretical spectra are in accord with experimental results presented in Ref.\cite{9}. Although the gold nanoshells have weaker absorption in the visible range compared to their single-component gold counterparts, the optical spectra of the nanocomposites are broaden since the presence of the additional optical resonance in near infrared region. Note that the infrared radiations carry 53 $\%$ of the total solar energy, while the visible region contributes to 43 $\%$ of the total solar heat flux. One could expect gold nanomatryoshkas are more suitable for solar cell applications.

\begin{figure}[h]
\includegraphics[width=8cm]{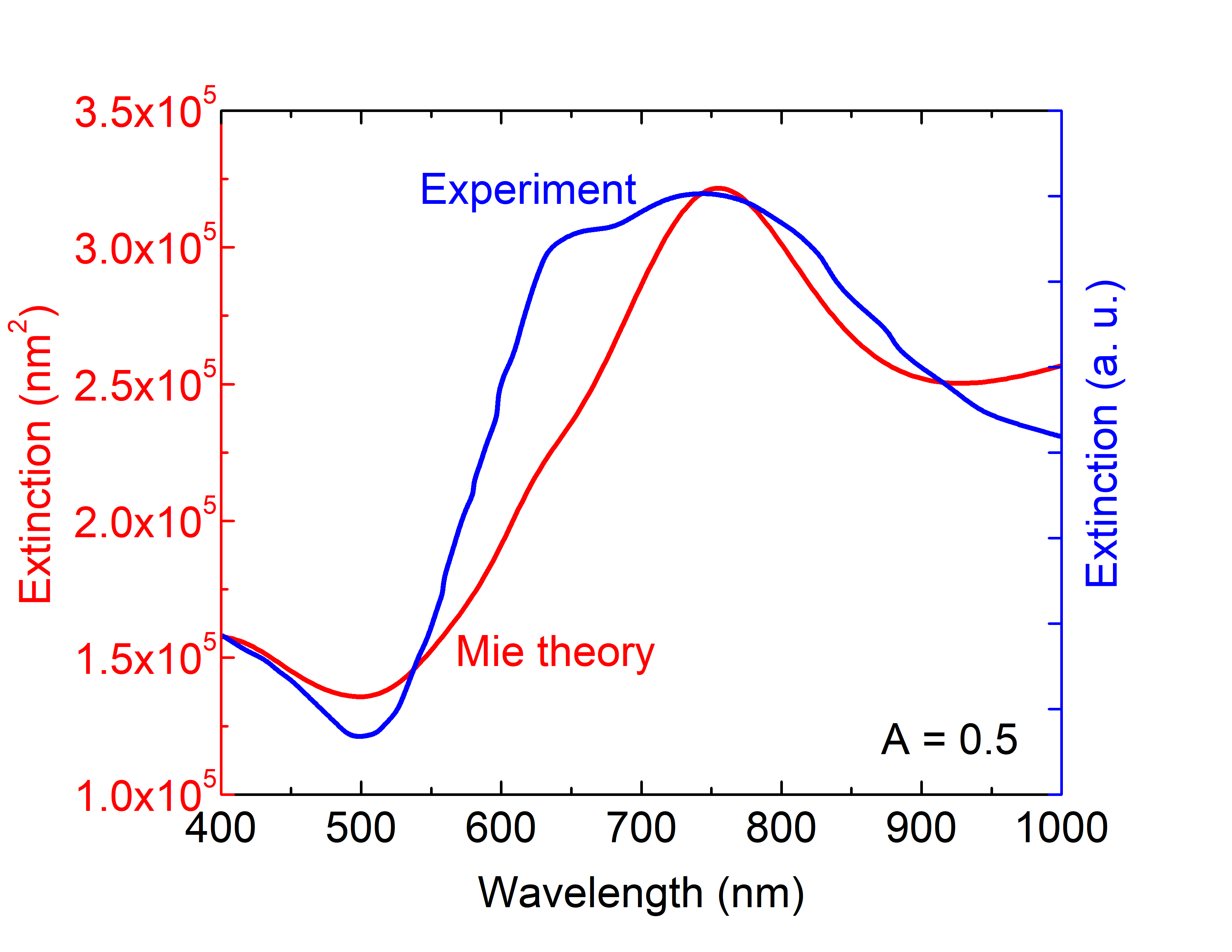}
\caption{\label{fig:3}(Color online) Theoretical extinction cross section of gold nanomatryoshkas of geometry [$r_1,r_2,r_3$]=[25,125,150] nm to describe experimental results presented in Ref.\cite{10}.}
\end{figure}

Figure \ref{fig:3} shows experimental ultraviolet-visible extinction spectra of a solution of gold nanoshells of size [$r_1,r_2,r_3$]=[$25,125,150$] nm in Ref.\cite{10} and our theoretical Mie calculations with $A = 0.5$ for the system. The theoretical prediction and experimental data are close to each other and show an optical resonance peak near 750 nm, but they do not perfectly overlap. This discrepancy is expected, since since Mie theory cannot capture all effects of surface roughness and morphology of the nanostructure in one parameter. However, we give a better agreement with experiment, compared with the FDTD simulation in Ref.\cite{10}. This finding is additional evidence for the importance of considering size-dependent dielectric functions in plasmonic properties of nanoscale materials. Additionally, the presence of only one peak in the spectrum suggests that the incoming electromagnetic field is shielded from the gold core by the dielectric layer. In other words, the plasmonic coupling between two metal layers is significantly weakened when the dielectric thickness is sufficiently large.

\begin{figure}[h]
\includegraphics[width=8cm]{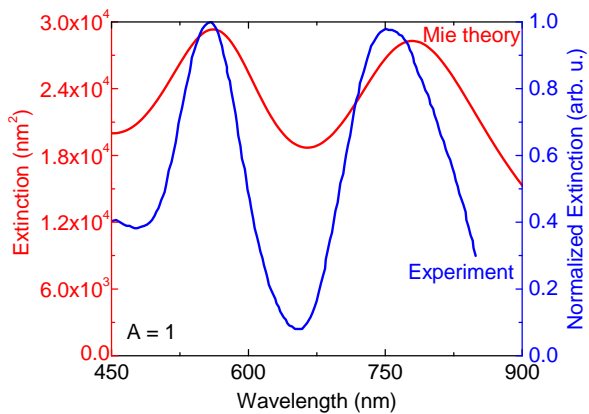}
\caption{\label{fig:4}(Color online) Theoretical (red) and experimental (blue) UV-vis extinction spectrum in Ref.\cite{6} obtained from aqueous suspensions of Au nanomatryoshkas having dimensions [$r_1,r_2,r_3$]=[25.5,38.4,53] nm.}
\end{figure}

Figure \ref{fig:4} presents extinction measurements for gold nanomatryoshkas of size [$r_1,r_2,r_3$] = [25.5,38.4,53] nm \cite{6} and our theoretical extinction cross section using Mie theory with $A = 1$. Our numerical results agree well with the experimental results. The authors in Ref. \cite{6} calculated theoretical extinction spectra also using Mie theory but associated with the dielectric functions for gold from John and Christy \cite{7}. Interestingly, the theoretical analysis in this prior work is fully consistent with the experiment. While one can expect the extinction intensity at the second peak is higher than that at the first peak as a consequence of ignoring size effects, similar to the numerical results in Fig.\ref{fig:2}a as setting $A = 0$. However, 27 separate Mie extinction spectra generated by varying the nominal dimensions along with their experimental errors were averaged to consider influences of the nanoshell size distribution on optical features and to depress the second extinction maximum. Consequently, both our approach and the previous estimate in Ref.\cite{6} confirm the effect of size on making accurate quantitative predictions.

Now one can use Mie theory with $A = 1$ to evaluate figure of merits (FoMs) for gold nanomatryoshkas to compare sunlight absorption properties between different materials and geometries. The figure of merit is defined as \cite{22}
\begin{eqnarray}
FoM = \int_{300}^{1400}\frac{Q_{abs}(\lambda)}{\pi r_3^2}\frac{E_{\lambda}}{E_{\lambda,max}}d\lambda,
\label{eq:12}
\end{eqnarray}
where $E_{\lambda,max}$ is the maximum of the spectral solar irradiance $E_\lambda$. 

\begin{figure}[h]
\includegraphics[width=8cm]{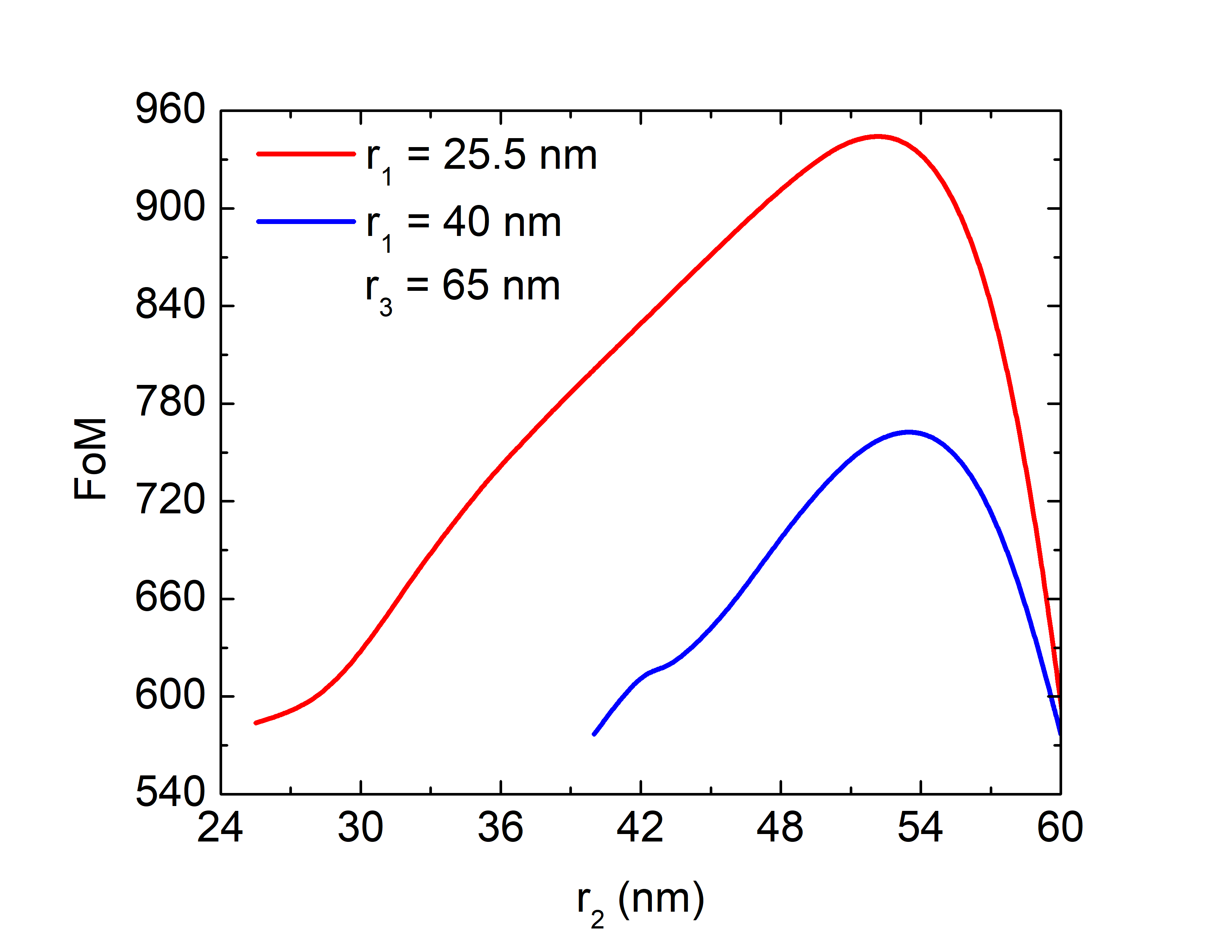}
\caption{\label{fig:9}(Color online) Figure of merit of three-layered composites having structure $\left[r_1,r_3\right] = [25.5,65]$ nm and $[40,65]$ nm as a function of $r_2$.}
\end{figure}

The figure of merit is strongly dependent on the size and dielectric function of spherical nanoparticles. The authors in Ref.\cite{31} reported that the FoMs of single component Au and TiN nanoparticles having radius of 50 nm are 657 and 683, respectively. The complex dielectric functions of the two materials in their Mie scattering calculations are based on density functional theory (DFT). While our predicted FoMs of the same size Au and TiN nanosphere are nearly 650 and 636, respectively. Here we extract the dielectric function data of a TiN film of thickness 180 nm from Ref.\cite{32}. Clearly, our predictions are relatively consistent with the DFT-based results. From this we extend our calculations to investigate the FoMs of 65-nm Au and TiN nanoparticles to compare with findings in previous work \cite{22} carried out by Ishii and his co-workers. Ishii reported that the Au and TiN nanoparticles having 65-nm radius possess the FoMs of 400 and 1020, respectively. The bulk dielectric functions in his calculations are taken from previous experimental data \cite{7,33}. However, we find the values of 583 for Au and 609 for TiN. The deviation may be a consequence of ignoring size effects. Our numerical results and the previous paper \cite{31} evince the nearly same efficiency of solar energy harvesting driven by optical properties of Au and TiN nanoparticles.

In Fig. \ref{fig:9}, the FoMs of gold nanomatryoshkas are calculated by choosing $\left[r_1,r_3\right] = [25.5,65]$ nm and $[40,65]$ nm, motivated by experiments reported in Ref.\cite{9,6}, and varying $r_2$. {\color{red}When $r_1 = r_2$}, the multilayered nanostructure becomes a pure gold nanoparticle of radius 65 nm. An increase of $r_2$ causes an additional plasmonic peak to emerge and the absorption spectrum to broaden. This effect is the result of collective oscillations of conduction electrons on metal surfaces, which couple together. Thus the FoM increases up to $\approx$ 950 as $r_2 =$ 52-54 nm for the system of $\left[r_1,r_3\right] = [25.5,65]$ nm. For the structure of $\left[r_1,r_3\right] = [40,65]$ nm, the maximum FoM is found to be $\approx$ 760 at $r_2 =$ 52-55 nm. Our findings suggest that one can optimize the figure of merit by manufacturing $r_3-r_2 \approx 10-12$ nm. When the thickness of the outermost gold shell is smaller than 10 nm ($r_3-r_2 \leq 10$ nm), the plasmonic decoupling between two metal-dielectric interfaces becomes greater. Consequently, the FoM is significantly reduced.


\begin{figure}[h]
\includegraphics[width=8cm]{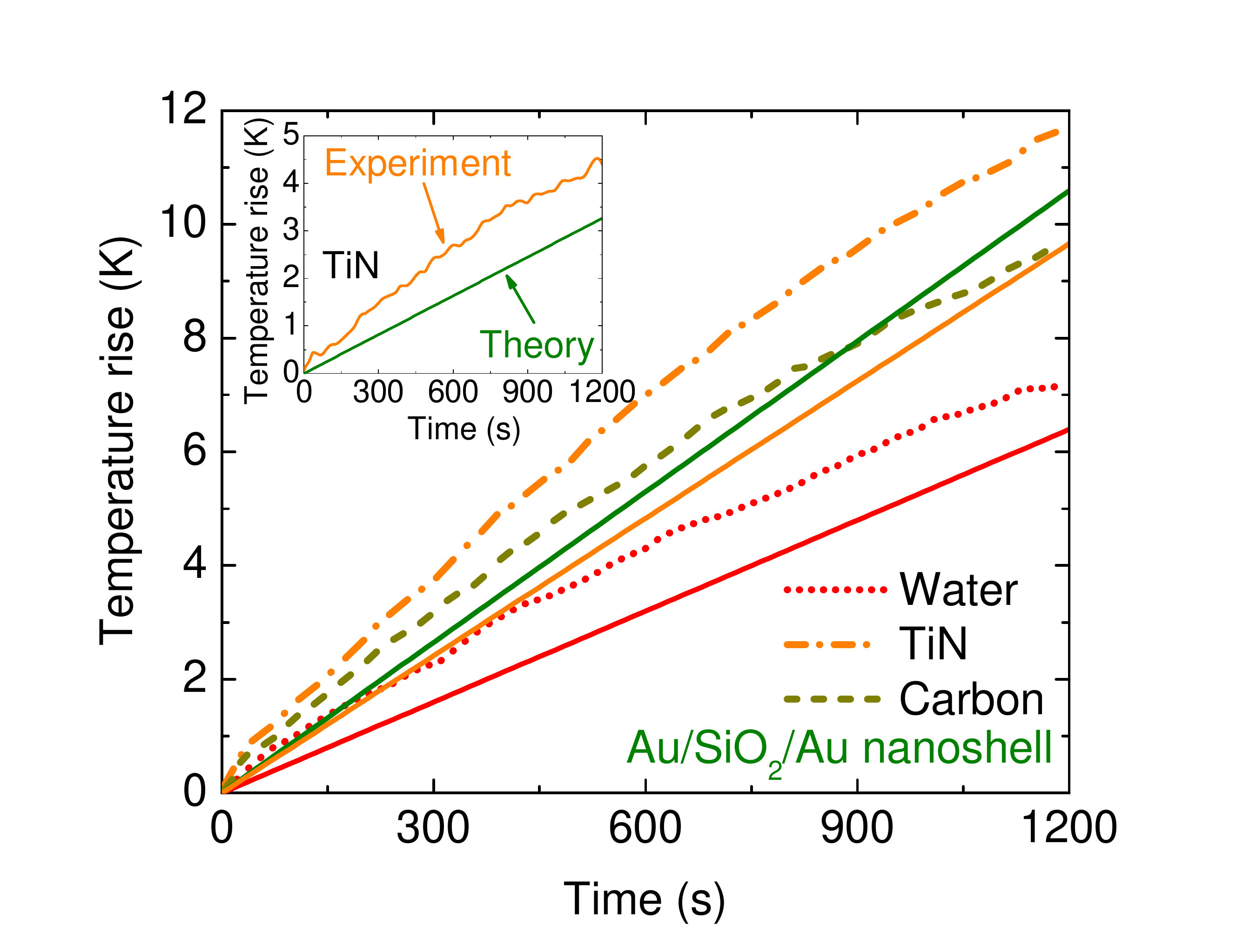}
\caption{\label{fig:8}(Color online) Theoretical temperature rise of water (red solid line), and a solution of gold nanomatryoshkas ({\color{red}dark} green solid line) with geometry [$r_1,r_2,r_3$]=[25.5,38.4,65] nm {\color{red}and a solution of 65-nm TiN nanoparticles (orange solid line)} as a function of time. Dotted, dashed and dashed-dotted lines correspond to experimental data \cite{22} of the temperature rise of water, TiN and carbon solution having a concentration of 0.0001 vol $\%$, respectively. {\color{red}Inset: Temperature increase as a function of time due to a contribution of TiN nanoparticles calculated using theory (dark green) and experiment (orange).}}
\end{figure}

The theoretical analysis of Eqs.(\ref{eq:10}), (\ref{eq:11}) and (\ref{eq:7}) can be applied to predict temperature rise of {\color{red} TiN nanospheres of radius 65 nm and} gold multilayered nanostructures having size of [$r_1,r_2,r_3$]=[25.5,38.4,65] nm illuminated by the solar simulator. Figure \ref{fig:8} shows that our theoretical predictions are close to experimental data in Ref.\cite{22} at the same concentration. {\color{red}As discussed above, using dielectric function data of TiN in different literatures generates various values of the FoM and optical absorption. One can think that a deviation between theory and experiment for the heating of TiN nanoparticle suspension is possibly attributed to a use of the TiN dielectric function extracted from experiments in Ref.\cite{32}. As can be seen in the inset of Fig. \ref{fig:8}, the contributions of TiN nanoparticle in the solution are simply calculated by subtracting the temperature increase in both theoretical calculations and experimental data of the solution with and without TiN nanoparticles. Two calculations are in accord with each other but the overlap is not perfect. Another reason has to be in our minimalist model of heat production in water although it appears the treatment can make a good prediction. Improving this calculation can perform better quantitative studies for the heating process.} {\color{red}In Ref. \cite{22}}, the greater FoM of a TiN nanosphere ($\approx$ 1020) than that of a carbon nanoparticle ($\approx$ 580) was found to be a main reason for the higher {\color{red}experimental} temperature change in the TiN solution compared with the carbon nanofluid. {\color{red}Similarly}, for the gold matryoshakas {\color{red} solution} at concentration of 0.0001 vol $\%$, an increase of temperature is larger but relatively close to the {\color{red}theoretical} solar thermal heating of the TiN solution because of the similar FoMs (FoM $\approx$ 770 for the gold nanoshell {\color{red}and FoM $\approx$ 609 for TiN in our calculations}). 

Our finding indicates that the FoM is a decisive factor for temperature rise caused by radiation. If our nanoparticle solution were exposed to real solar irradiation, we could unify limits in the integrals of Eqs. (\ref{eq:10}) and (\ref{eq:12}) and recast Eq.(\ref{eq:7}) by
\begin{eqnarray}
\rho c V_0\frac{\partial \Delta T}{\partial t} = N\pi r_3^2E_{\lambda,max} V_0FoM + P_{abs}^{w}4\pi R^2.
\label{eq:13}
\end{eqnarray}
The physical assumption underlying Eq.(\ref{eq:13}) does not capture a vaporization process. It works relatively well for a solution concentration less than 0.0001 vol $\%$. One can expect a nanocomposite density increase causes the temperature increase in solution. However, at higher concentrations, the time-dependent temperature variations have been found to stop increasing \cite{22}. The behavior is attributed to solar vapor/steam generation. This problem is under study. 

\section{Conclusions}
We have shown that gold nanomatryoshkas are advantageous for solar-cell applications because of their stronger plasmon resonances and broader plasmon bands, compared with their single-component gold counterparts. Titanium nitride nanoparticles was found to generate more solar-to-thermal energy than solid gold nanospheres \cite{22}. However, our results reveal that a silica dielectric layer sandwiched between two gold layers significantly enhances the light trapping efficiency of gold-based nanocomposites. The efficiency enhancement of solar energy harvesting is highly tunable. The figure of merit of gold nanomatryoshkas having appropriate structures can be larger than that of the same size TiN nanoparticles. We have proposed a simple calculation to estimate the time-dependent temperature of the Au/\ce{SiO_2}/Au nanoshell solutions under simulated solar illumination. The prediction cannot be accurate if structural and size effects are ignored. Our theoretical modeling agrees well with experiments in some limits \cite{22}. The approach would enable us to step forward in understanding surface plasmon resonance behaviors of arbitrary nanostructures to optimize sunlight absorption.

\begin{acknowledgments}
This research is funded by Vietnam National Foundation for Science and Technology Development (NAFOSTED) under grant number 103.01-2017.63. K.W. acknowledges the financial support by JSPS KAKENHI Grants No. 18H01154. 
\end{acknowledgments}

\end{document}